\documentclass[aip,onecolumn, groupedaddress,showpacs,floatfix]{revtex4-1}

\usepackage{dcolumn}
\usepackage{amsmath}
\usepackage{amssymb}
\usepackage{graphicx}
\usepackage{bm}
\usepackage[T1]{fontenc}
\usepackage{color}
\usepackage{url}
\usepackage{rotating}
\usepackage{todonotes}

\usepackage{wasysym}
\usepackage[latin1]{inputenc}
\usepackage[T1]{fontenc}
\usepackage{ae,aecompl}

    \DeclareGraphicsExtensions{.pdf}

\newcommand{\abs}[1]{\vert#1\vert}
\newcommand{\be}{\begin{equation}}
\newcommand{\ee}{\end{equation}}
\newcommand{\bea}{\begin{eqnarray}}
\newcommand{\eea}{\end{eqnarray}}

\newcommand{\eye}{\mbox{$\mbox{1}\!\mbox{l}\;$}}

\DeclareMathOperator{\Real}{Re}

\begin{document}

\title{Taming Instabilities in Power Grid Networks by Decentralized Control}

\author{Benjamin Sch\"afer}
\affiliation{Network Dynamics, Max Planck Institute for Dynamics and Self-Organization (MPIDS), 37077 G\"ottingen, Germany}
\author{Carsten Grabow}
\affiliation{Potsdam Institute for Climate Impact Research, 14412 Potsdam, Germany}
\author{Sabine Auer}
\affiliation{Potsdam Institute for Climate Impact Research, 14412 Potsdam, Germany}
\affiliation{Department of Physics, Humboldt University Berlin, 12489 Berlin, Germany}
\author{J\"urgen Kurths}
\affiliation{Potsdam Institute for Climate Impact Research, 14412 Potsdam, Germany}
\affiliation{Department of Physics, Humboldt University Berlin, 12489 Berlin, Germany}
\affiliation{Institute of Complex Systems and Mathematical Biology, University of Aberdeen, Aberdeen AB24 3FX, UK}
\affiliation{Department of Control Theory, Nizhny Novgorod State University, 606950 Nizhny Novgorod, Russia}
\author{Dirk Witthaut}
\affiliation{Forschungszentrum J\"ulich, Institute for Energy and Climate Research (IEK-Systems Analysis and Technology Evaluation)}
\affiliation{Institute for Theoretical Physics, University of Cologne, 50937 K\"oln, Germany}
\author{Marc Timme}
\affiliation{Network Dynamics, Max Planck Institute for Dynamics and Self-Organization (MPIDS), 37077 G\"ottingen, Germany}
\affiliation{Institute for Nonlinear Dynamics, Faculty of Physics, University of G\"ottingen, 
37077 G\"ottingen, Germany}

\date{\today}

\begin{abstract}
	Renewables will soon dominate energy production in our electric power system. 
	And yet, how to integrate renewable energy into the grid and the market is still a subject of major debate.
	Decentral Smart Grid Control (DSGC) was recently proposed as a robust and decentralized approach to balance supply and demand and to guarantee a grid operation that is both economically and dynamically feasible.                        
	Here, we analyze the impact of network topology by assessing the stability of essential network motifs using both linear stability analysis and basin volume for delay systems. Our results indicate that if frequency measurements are averaged over sufficiently large time intervals, DSGC enhances the stability of extended power grid systems. We further investigate whether DSGC supports centralized and/or decentralized power production and find it to be applicable to both. However, our results on cycle-like systems suggest that DSGC favors systems with decentralized production. Here, lower line capacities and lower averaging times are required compared to those with centralized production.
\end{abstract}


\maketitle

\section{Introduction}
\label{Introduction}
The ongoing climate change is forcing us to shift our power generation from fossil power plants towards renewable generation \cite{mitigation2011ipcc}. In the last years, renewable energy technology development and policy support led to a tremendous increase in the share of Renewable Energy Sources (RES). In 2014, Germany covered $27.8 \%$ of its gross electricity consumption with RES \cite{EE2014}.  Still, large conventional power plants dominate the power grids: transmission lines connect large plants with regional consumers in a locally star-like topology. With more renewable power sources entering, the grid topologies become more decentralized and more recurrent due to the distributed generation \cite{ackermann2001distributed}. In such a scenario, consumers may act as producers and consumers at the same time, so-called \emph{prosumers} \cite{kotler1986prosumer} and electricity transport is no longer unidirectional. 

A known challenge of many renewable sources is their volatile nature \cite{Turner1999,50Hertz2012,Boyle2004}. Fluctuations occur on different time scales, including seasonal, inter-day \cite{Heide2010} and intra-second fluctuations \cite{Milan2013}. This requires radical changes in the control and design strategies  of electric power grids as well as market innovations to ensure cost effectiveness.
Therefore, a need for more flexibility options for power market supply and balancing energy \cite{gruenbuch} arises because the fluctuating RES cannot guarantee power supply with the certainty conventional plants could.  In this regard, it is most important to identify options that are both cost efficient and system stabilizing. So far, the framework of power market design and power grid stability with its long planning horizons does not satisfy the need for sufficient flexibility options \cite{BEEStudie}. 

Different \emph{smart grid} approaches have been proposed to present ways to match supply and demand in such a fluctuating power grid. However, economic and political feasibility and market integration are often missed out. A key idea of various smart grid concepts is to regulate the consumers' demand  \cite{Butler2007}, a massive paradigm shift compared to the current power grid operation schemes \cite{Albadi2008,Palensky2011}. Many proposals for smart grids are based on sufficient information and communication technology infrastructure, see, e.g.,~\cite{Kok2005} or \cite{Hofmann}. 
However, such a centralized system would raise questions of cyber security and privacy protection \cite{Ericsson2010,Fang2012} and several studies highlight the cost burden these proposals implicate \cite{meterstudy}.

In contrast, an alternative approach without massive communication between consumers and producers directly utilizes the grid frequency to adjust production and consumption. The frequency increases in times of power excess while it decreases in times of underproduction \cite{Schweppe1982,Short2007}.
A novel smart grid concept, Decentral Smart Grid Control (DSGC), was introduced in \cite{Walter}, based on earlier ideas by \cite{Schweppe1982}, and its mathematical model proposed and analyzed in \cite{Schaefer2015}.
Using DSGC prosumers control their momentary demand on the basis of the grid frequency which can easily be measured everywhere with cheap equipment.
Ref. \cite{Schaefer2015} demonstrates that DSGC enhances the stability of the power grid dynamics if the frequency measurements are averaged over sufficiently long time intervals. Yet, so far, only very small networks were investigated. Hence, the impact of grid topologies on power grid dynamics with DSGC constitutes a widely open research question.

Here we analyze the impact of network topology by assessing the  stability of essential network motifs using both linear stability  analysis of delay systems and determining
basin volume. Furthermore, we address the question, how grid stability changes when generation is decentralized. The article is structured as follows. First, we present a dynamical model for power grid dynamics
and present the concept of Decentral Smart Grid Control (DSGC) \cite{Schaefer2015} to control a power grid in section \ref{Model}.
In section \ref{Stability_methods}, we briefly summarize linear stability and basin volume measures for such delayed systems. The stability results of DSGC are then presented for a star motif in section \ref{Stability_results_4_node} where we discuss the destabilizing resonance and rebound effects and how stable grid operation remains possible. Using linear stability analysis, we investigate the effect of decentralized power generation
in cyclic and square lattice grid motifs in section \ref{Decentralized_production}. The results suggest that DSGC works successfully for centralized as well as decentralized production, where grids with decentralized production require lower line capacities than centralized ones.

\section{Coupled oscillator model with Decentral Smart Grid Control}
\label{Model}
To model the frequency dynamics of a large-scale power grid, we consider an oscillator model based on the physics of coupled synchronous generators and synchronous motors, see \cite{Filatrella2008,Witthaut2012,Rohden2014,Doerfler2013,Motter2013,Menck2014,Rohden2012} for details.
This model is similar to the "classical model" \cite{Machowski2011}
and the "structure preserving model" \cite{Bergen1981} from power
engineering.

The state of each machine $i\in \{1,...,N\}$ is characterized by the rotor angle
$\theta_i(t)$ relative to the grid reference rotating at $\Omega = 2\pi \times
50$ Hz or $\Omega = 2 \pi \times 60$ Hz, respectively, and its angular frequency deviation
$\omega_i=\mathrm{d} \theta_i / \mathrm{d}t$  from the reference. 
Each machine is driven by a mechanical power $P_i(t)$, which is positive for a generator and negative for a consumer. In addition, every machine transmits electric power via the adjacent transmission lines which have a	coupling strength $K_{ij}$. This coupling strength expresses the maximal possible power that may theoretically be transmitted through the power lines. The dynamics of the machine $i$ is then given by the equation of motion as

\begin{equation}
\label{eq.motion}
\frac{\mathrm{d}^2\theta_i}{\mathrm{d}t^2}=P_i-\alpha_i\frac{\mathrm{d}\theta_i}{\mathrm{d}t}+\sum_{j=1}^N K_{ij}\sin(\theta_j-\theta_i)
\quad \forall i \in \{1,...,N\},
\end{equation}
where $\alpha_i$ is a damping constant. We neglect ohmic loads which should be small compared to shunt admittances \cite{VanHertem2006} for the dynamics we consider. 
We take the moment of inertia to be identical for all machines and hence eliminate such moments of inertia in the equation of motion for simplicity of presentation.
Equation (\ref{eq.motion}) as well as the upcoming equations (\ref{eq.motion with delay}) and (\ref{eq.motion with average}) are discussed in more detail in \cite{Schaefer2015}.

Decentral Smart Grid Control (DSGC) is based on Demand Response that aims to stabilize the power system by encouraging consumers to lower their consumption in times of high load and low production and increase consumption in times of low load but high production. 
Instead of paying a constant price for electric power, consumers are presented with a linear price-frequency relation $p_i(\frac{\mathrm{d}\theta_i}{\mathrm{d}t})$ 
\begin{equation}
\label{eq:linear_price}
p_i(\frac{\mathrm{d}\theta_i}{\mathrm{d}t})=p_\Omega-c_1\cdot \frac{\mathrm{d}\theta_i}{\mathrm{d}t}
\end{equation}
to motivate grid-stabilizing behavior. Although consumer reaction might be very complex, we assume a linearized power-price relation $\hat{P_i}(p_i)$ 
\begin{equation}
\label{eq:linear_power}
\hat{P_i}(p_i)\approx P_i+c_2\cdot (p_i-p_\Omega)
\end{equation} by the consumers close to the stable operational state. Plugging (\ref{eq:linear_price}) into (\ref{eq:linear_power}) and defining $\gamma=c_1\cdot c_2$ leads to a linear response of consumed and produced mechanical power $\hat{P_i}(t)$ as a function of frequency deviation $\mathrm{d}\theta_i/\mathrm{d}t$:

\begin{figure}[hbt!]
	\centering
	\includegraphics[width=14cm]{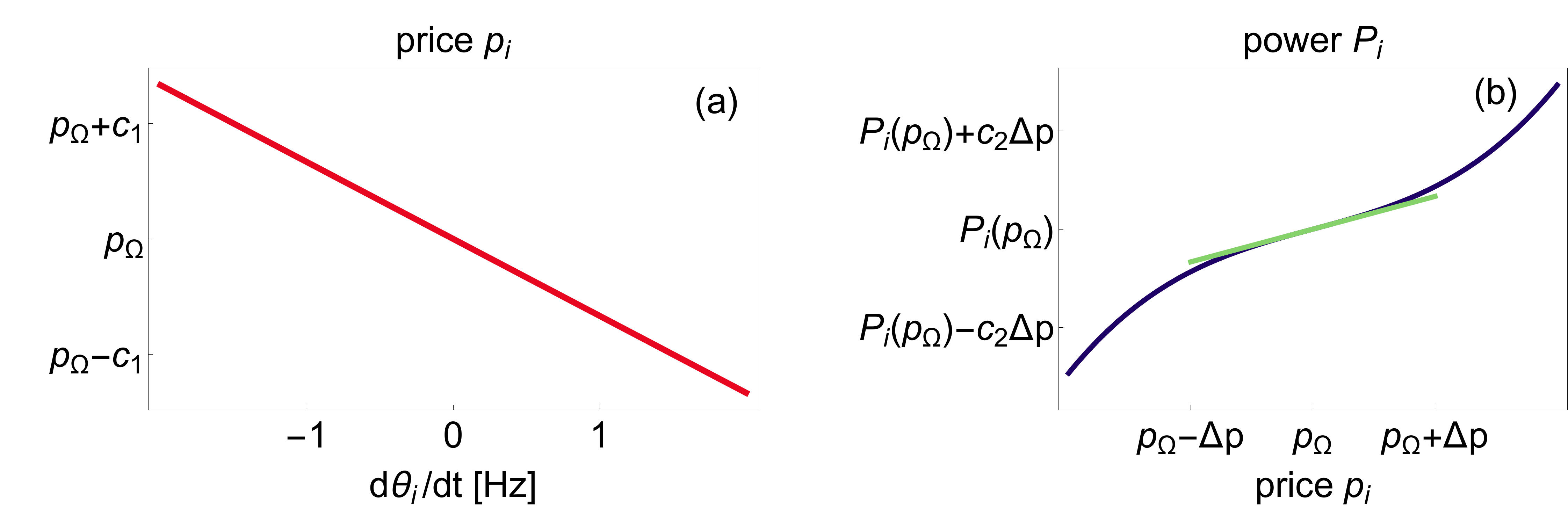}
	\caption{\label{fig:price_frequency_power}
		\textbf{Using linear relations the power becomes a linear function of the frequency deviation $\mathrm{d}\theta_i/\mathrm{d}t$.} (a): We assume a linear price-frequency relation to motivate consumers to stabilize the grid. For example, if the production is larger than consumption, the power grid frequency increases. Hence, decreasing prices should motivate additional consumption. (b): Although consumers might react non-linearly towards price-changes (dark blue), we assume a linear relationship (light green) close to the operational frequency $\Omega$ which corresponds to $\mathrm{d}\theta_i/\mathrm{d}t=0$.}
\end{figure}

\begin{equation}
\label{power-frequency-relation}
\hat{P_i}(t)=P_{i}-\gamma_i\frac{\mathrm{d}\theta_i}{\mathrm{d}t}(t)
\quad \forall i \in \{1,...,N\},
\end{equation}
where $\gamma_i$ is proportional to the price elasticity of each node $i$, i.e., measures how much a producer or consumer is willing to adapt their consumption or production, see also \cite{Schaefer2015}.
In general, such an adaptation will not be instantaneous but will be delayed by a certain time $\tau$ by a measurement and the following reaction. 
We can now substitute the function $\hat{P_i}(t-\tau)$ from (\ref{power-frequency-relation}) for the fixed value $P_i$ in the uncontrolled system (\ref{eq.motion}) and obtain the equation of motion 

\begin{equation}
\label{eq.motion with delay}
\frac{\mathrm{d}^2\theta_i}{\mathrm{d}t^2}=P_i-\alpha_i\frac{\mathrm{d}\theta_i}{\mathrm{d}t}+\sum_{j=1}^N K_{ij}\sin(\theta_j-\theta_i)-\gamma_i\frac{\mathrm{d}\theta_i}{\mathrm{d}t}(t-\tau)
\quad \forall i \in \{1,...,N\},
\end{equation}
with DSGC including a delayed power adaptation.
In \cite{Schaefer2015} it was already shown that such a delayed system poses risks to the stability of the power grid for certain delays $\tau$. 
Hence, an extension using frequency measurements averaged over time intervals of lengths $T$ were introduced to stabilize the power grid regardless of the specific delay. Such averaging yields

\begin{eqnarray}
\label{eq.motion with average}
\frac{\mathrm{d}^2\theta_i}{\mathrm{d}t^2} & = &P_i-\alpha_i\frac{\mathrm{d}\theta_i}{\mathrm{d}t}+\sum_{j=1}^N K_{ij}\sin(\theta_j-\theta_i)-\frac{\gamma_i}{T}\int_{t-T}^{t}\frac{\mathrm{d}\theta_i}{\mathrm{d}t}(t'-\tau) \mathrm{d}t'\\
& = & P_i-\alpha_i\frac{\mathrm{d}\theta_i}{\mathrm{d}t}+\sum_{j=1}^N K_{ij}\sin(\theta_j-\theta_i)-\frac{\gamma_i}{T} \left(\theta_i(t-\tau)- \theta_i(t-\tau-T) \right)
\quad \forall i \in \{1,...,N\}.
\end{eqnarray}
For what follows, we choose homogeneous averaging time $T$ for all machines, as well as similar delays $\tau$ for all nodes. In addition, we use homogeneous capacities $K_{ij}=K$ for all lines to simplify the calculations. 
In the section \ref{results} we apply equations (\ref{eq.motion with delay}) and (\ref{eq.motion with average}) to different network topologies and evaluate their stability as a function of the delay $\tau$ with different averaging times $T$. We hereby treat the averaging time $T$ as a control parameter that can be chosen when setting up the system, while the delay $\tau$ remains as an exogenous parameter introduced by the consumers and producers.

\section{Assessing robust operation}
\label{Stability_methods}
Here, we discuss how linear stability analysis and measuring basin volume yield information about robust operation, in dependence of delay $\tau$ and for different averaging times $T$. 
First, we introduce the fixed point of the system, then discuss linear stability analysis of delayed systems and finally point out difficulties when assessing basin volume of a power grid system with delay.

\paragraph{Fixed points.}
To study the stability and the role of the system parameters,
we analyze the 
dynamical stability around the steady-state operation of the grid given by the fixed point  
\begin{equation}   
\left( \theta_i(t),\frac{\text{d}\theta_i}{\text{d}t}(t)\right)=\left( \theta^{*}_i,\omega^{*}_i\right)
\quad \forall i \in \{1,...,N\},
\label{eqn:osc-fixedpoint}
\end{equation}
as obtained by solving
\begin{equation}   
 \frac{\mathrm{d^2}}{\mathrm{d}t^2} \theta_i=\frac{\mathrm{d}}{\mathrm{d}t} \theta_i=0
 \quad \forall i \in \{1,...,N\}.
\label{eqn:osc-fixedpointDetermination}
\end{equation}
Since $\omega_i=\mathrm{d}\theta_i/\mathrm{d}t $ we directly obtain $\omega^{*}_i=0$ for all $i$. Hence, we only need to determine the angles $\theta^{*}_i$.
A fixed point can only exist, if the grid has a sufficient transmission capacity $K_{ij}$ to transmit the power from the producers to the consumers \cite{Manik2014}. The minimal $K_{ij}$ for which a stable fixed point exists is called critical coupling \cite{Rohden2014}.

\paragraph{Linear stability.}
Linear stability of a dynamical system is determined by the eigenvalues of
its characteristic equation. For systems without delay this is a polynomial obtained from the Jacobian of the system but for a delayed system it becomes a quasi-polynomial with infinitely many solutions \cite{Driver1977,Roussel2004}. We obtain the characteristic equation by calculating the Jacobian of both the non-delayed system, based on equation  (\ref{eq.motion with delay}) with $\tau =0$,
\begin{equation}
J_{0}=\left(\begin{array}{cc}
\frac{\partial}{\partial \theta_i}\left(\frac{\mathrm{d}}{\mathrm{d}t} \theta_j\right)
&
\frac{\partial}{\partial\omega_i}\left(\frac{\mathrm{d}}{\mathrm{d}t}
\theta_j\right)\\
\frac{\partial}{\partial \theta_i}\left(\frac{\mathrm{d}}{\mathrm{d}t}
\omega_j\right) &
\frac{\partial}{\partial\omega_i}\left(\frac{\mathrm{d}}{\mathrm{d}t}
\omega_j\right)
\end{array}\right) \in \mathbb{R}^{2N\times 2N},
\label{eqn:delay-J0}
\end{equation}
and the derivatives for the delayed terms involving $\tau$,
\begin{equation}
J_{\tau}=\left(\begin{array}{cc}
\frac{\partial}{\partial \theta_{\tau , i}}\left(\frac{\mathrm{d}}{\mathrm{d}t} \theta_j\right)
&
\frac{\partial}{\partial\omega_{\tau , i}}\left(\frac{\mathrm{d}}{\mathrm{d}t}
\theta_j\right)\\
\frac{\partial}{\partial \theta_{\tau ,  i}}\left(\frac{\mathrm{d}}{\mathrm{d}t}
\omega_j\right) &
\frac{\partial}{\partial\omega_{\tau , i}}\left(\frac{\mathrm{d}}{\mathrm{d}t}
\omega_j\right)
\end{array}\right) \in \mathbb{R}^{2N\times 2N},
\label{eqn:delay-Jtau}
\end{equation}
where we abbreviated $\theta_{\tau ,i}=\theta_i(t-\tau)$ and $\omega_{\tau ,i}=\frac{\mathrm{d\theta_i}}{\mathrm{d}t}(t-\tau)$ and $i,j \in \{1,...,N\}$. We hereby consider exponentially decaying or growing solutions \cite{Driver1977}.
The stability eigenvalues $\lambda$ are then determined by the solutions of the characteristic equation
\begin{equation}
p(\lambda)=\det( J_0 + {\rm e}^{- \lambda \tau} J_\tau - \lambda \eye)=0.
\label{eq:characteristic equation for delay}
\end{equation}
For the delayed system with averaging, i.e., equation (\ref{eq.motion with average}), we simply calculate the delayed Jacobian for the  two delays $\tau$ and $\tilde{T}=T+\tau$. Hence, the characteristic equation is given by 
\begin{equation}
p(\lambda)=\det( J_0 + {\rm e}^{- \lambda \tau} J_\tau + {\rm e}^{- \lambda \tilde{T}} J_{\tilde{T}}- \lambda \eye)=0.
\label{eq:characteristic equation for averaging}
\end{equation}
We obtain the symbolic expression for the characteristic equation using Mathematica \cite{WolframResearch2014} which is then also used to numerically determine roots of the characteristic equation, via Newton's method. Equations (\ref{eq:characteristic equation for delay}) and (\ref{eq:characteristic equation for averaging}) have infinitely many solutions but only a finite number of those can have a positive real part and those determine the instability of the system \cite{Gu2003}. 

Our method of finding these eigenvalues works as follows: We start at an arbitrary delay $\tau =\tau_\text{sampling}>0$ and let Mathematica find approximately 10,000 roots by choosing random complex initial conditions for Newton's algorithm. Afterwards, we delete double entries. The obtained eigenvalues are taken as the initial conditions for Newton's algorithm for the next larger delay $\tau=\tau_\text{sampling}+0.01$s. These eigenvalues then serve as the initial conditions for the next delay step etc. Similarly, we obtain eigenvalues for smaller delays like $\tau=\tau_\text{sampling}-0.01$s by using again the eigenvalues from $\tau_\text{sampling}$ as initial conditions.

Linear stability analysis quantifies whether a fixed point is stable to small perturbations and constitutes a fundamental aspect of stability in dynamical system. Assessing the stability of the system with respect to larger perturbations requires further analysis.

\paragraph{Basin volume.}
The global stability of a fixed point of a dynamical system can be quantified by the volume of its basin of attraction. 
An estimate for the basin volume $V_\text{basin}$ is determined numerically using a Monte Carlo method as the ratio of initial conditions converging to a stable operation state to the total number of initial conditions, as proposed in \cite{Menck2013}. Note that delayed systems are infinite-dimensional \cite{Driver1977} and do need an initial function instead of a single initial condition. We treat this problem by setting the initial function to be identical to the initial condition for all times smaller than zero, i.e,

\begin{equation}
\theta(t\leq 0)=\theta(t=0),\quad \omega(t\leq 0)=\omega(t=0).
\end{equation}
Thereby, we can effectively choose initial conditions as they completely define the initial function. In the following we take $M=1000$ randomly chosen initial conditions into account in order to estimate the basin volume's dependence on the delay time $\tau$. 

We are mainly interested in how fluctuations or disturbances in the energy generation will influence the system's dynamics. Hence, we first perturb the producer's node phase angle and angular velocity around its component of the fixed point (see fig. \ref{fig:4_node_eigenvalues}a for the network topology). In the next series of simulations, we perturb one randomly selected customer node around its component of the fixed point. Perturbations are uniformly chosen at random from the intervals $\Delta \theta_i \in [-\pi,\pi ]$ and $\Delta \omega_i \in [-30,30]$ Hz for the initial angles and initial frequencies respectively, similar to \cite{Menck2014,schultz2014detours}. We run the simulations for a simulation time of $t_\text{sim}=1500 \text{s}$. These long and computationally costly simulation times are necessary because we observed that for specific values of delay time $\tau$, e.g., $\tau= 1.4 \text{s}$, perturbations may decay relatively quickly toward the fixed point but later still escalate.

Note that we only consider so-called single node basin volumes, i.e., we only perturb the component the a fixed point of one node. In theory, all nodes could be perturbed simultaneously which results in a more complete sample of the phase space. Unfortunately, the total phase space volume grows exponentially with the number of nodes, making it infeasible to sample the full phase space.

\section{Results}
\label{results}

We now present results about networks with Decentral Smart Grid Control. First, we present and compare the results of linear stability and basin volume analysis of a four node star motif. This motif constitutes one of the main building blocks of power grids, since, in principle, its effective topology locally resembles a star, the central node being a large power plant that supplies the regional consumers in its vicinity \cite{50Hertz2012,Rohden2014}.  Hereby, we discuss the destabilizing effects of resonances and the "rebound effect" for large delays. Using basin volume we present how intermediate delays $\tau$ benefit the stability. Finally, we consider larger networks and demonstrate how decentralization enhances stability. The parameters of the swing equation are calculated from standard literature values \cite{Machowski2011,schultz2014detours}. 
In current (European) power plants the initial delays have to be smaller than 2 seconds according to European regulations \cite{ENTSO-E2013}, in practice they will be significantly smaller. However, in future power grids additional communication delays \cite{Naduvathuparambil2002} of the order of several hundred milliseconds might arise in addition to unknown delays caused by demand response and additional power electronics. Hence, we consider a large range of potential delays $\tau \in (0,5)$s looking for the boundary of acceptable delays.

\subsection{Stability of the star motif}
\label{Stability_results_4_node}

\begin{figure}[htb]
	\centering
	\includegraphics[width=0.95\linewidth]{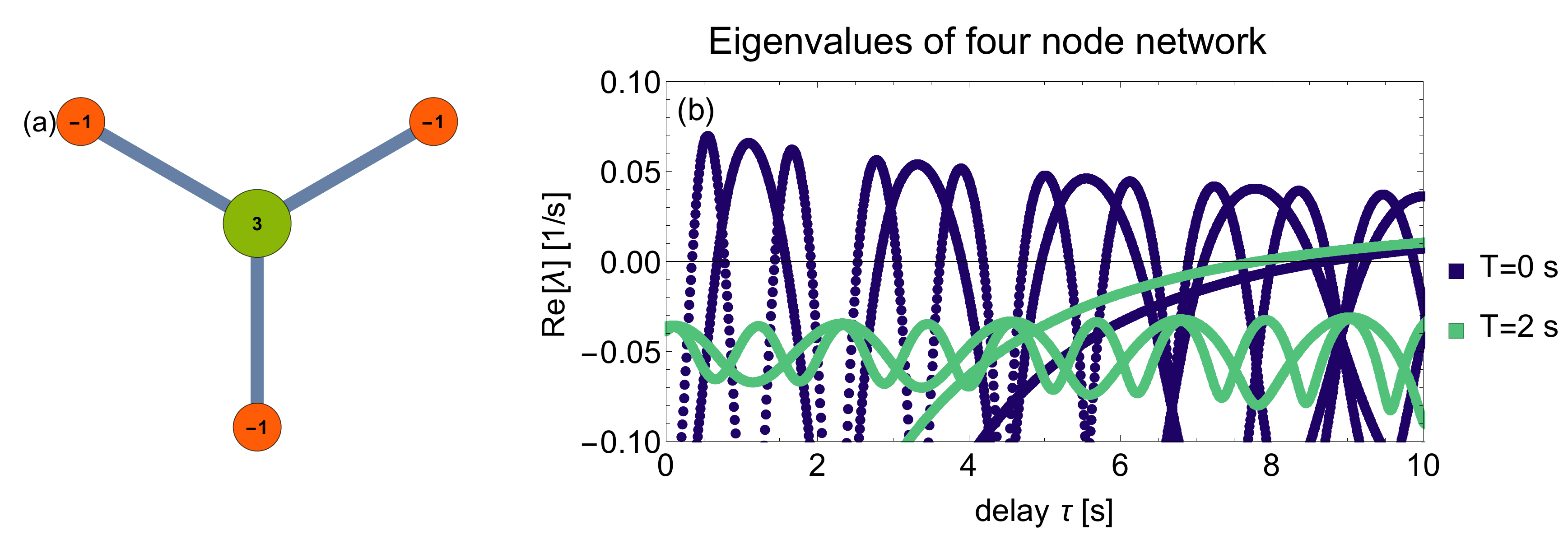}
	\caption{\label{fig:4_node_eigenvalues}
	\textbf{Resonances and large delays $\tau$ destabilize the four node system.} 
	(a): An elementary building block in a power grid with centralized production is shaped like a star. Shown is a motif whose linear stability and basin volume we study. The network is formed of one producer (green) in the center with power $P_\text{producer}=3/\text{s}^2$  and three consumers (red) with power $P_\text{consumer}=-1/\text{s}^2$ each.
	(b): Plotted are the eigenvalues with the largest real part as functions of delay $\tau$. For no averaging (dark blue curve), stable and unstable regions exist. For an averaging of $T=2\text{s}$, the system is stable for all delays below a critical $\tau_c\approx 8\text{s}$. In \eqref{eq.motion with average} parameters $\alpha =0.1/\text{s}$, $K=8/\text{s}^2$ and $\gamma =0.25/\text{s}$ were applied.}
\end{figure}

\paragraph{Delays induce destabilizing resonances.}
Networks with star topology (fig \ref{fig:4_node_eigenvalues}a) exhibit stability properties that depend crucially on the delay and the averaging applied (fig. \ref{fig:4_node_eigenvalues}b).
Without any averaging (fig. \ref{fig:4_node_eigenvalues}b dark blue curve), there are delays $\tau$ for which the fixed point is linearly unstable, i.e., there are eigenvalues with a positive real part $\Real{\lambda} \ge 0$. Those eigenvalues exhibit a periodic behavior with respect to the delay $\tau$. Operating the power grid at a delay $\tau$ for which we find a positive real part, e.g., $\tau \approx 1$s, is equivalent to resonantly driving the power grid away from the fixed point instead of damping it towards stable operation. These destabilizing delays are linked to the eigenfrequency of the oscillators in the power grid. If the delay is half the eigenoscillation duration, then it increases amplitudes of perturbations instead of damping them. This destabilization only occurs for $\alpha<\gamma$ because the resonant driving has to be larger than the intrinsic damping of the system, see also \cite{Schaefer2015}.
Introducing sufficiently large averaging times into the control cures these instabilities (fig. \ref{fig:4_node_eigenvalues}b light green curve); the unstable regions vanish for all delays $\tau < 7$s. 

\paragraph{Rebound effect for large delays.}
\label{Effect_of_large_delays}
For delays larger than a critical delay $\tau> \tau_c\approx 8.7\text{s}$ the system always gets destabilized, i.e., there is an eigenvalue with $\text{Re}(\lambda)>0$.
This rebound effect acts on a longer time scale than the intrinsic oscillations of the power grid system and originates from an over-reaction of the attempted damping as we explain below.
The existence of such a rebound effect is independent of averaging $T$ (fig. \ref{fig:4_node_eigenvalues}). We determine the critical delay without averaging $\tau_c$ to be

\begin{equation}
\label{eq:critical delay without average}
\tau_{c}=\frac{\arccos\left(-\frac{\alpha}{\gamma}\right)}{\sqrt{\gamma^{2}-\alpha^{2}}}+\frac{2\pi n}{\sqrt{\gamma^{2}-\alpha^{2}}},\,\, n\in\mathbb{Z}.
\end{equation}
This result is obtained by the following considerations. We define the sum of all angles as $\Sigma\theta:=\sum_{i=1}^{N}\theta_{i}$ and obtain its equation of motion by using eq. (\ref{eq.motion with delay}) as 
\begin{equation}
\frac{\mathrm{d^2}}{\mathrm{d}t^2} \Sigma\theta (t)= -\alpha\frac{\mathrm{d}}{\mathrm{d}t} \Sigma\theta (t) 
-\gamma\frac{\mathrm{d}}{\mathrm{d}t} \Sigma\theta (t-\tau).
\end{equation}
The characteristic equation of this equation reads
\begin{equation}
p\left(\lambda\right)=-\alpha-\gamma e^{-\lambda\tau}-\lambda=0,
\label{eq:simplified characteristic equation of sum}
\end{equation}
where we eliminated a zero eigenvalue $\lambda=0$ which arises due to the possibility to shift all angles by a constant.
For $\tau=0$ the eigenvalue $\lambda=-\alpha-\gamma$ is negative
as $\alpha>0$ and $\gamma>0$ and hence the system is stable with
respect to the sum $\Sigma\theta$. For larger delays $\tau>0$ we set
$\lambda=i\cdot\xi$ to obtain the delays for which the stability
changes. We get
\begin{equation}
-\alpha-\gamma e^{-i\xi\tau}-i\xi=0.
\end{equation}
Applying complex expansion and separating into real and imaginary
parts we obtain
\begin{eqnarray}
-\alpha-\gamma\cos(\xi\tau) & = & 0
\label{eq:real part of characteristic equation at edge of stability}\\
\gamma\sin(\xi\tau)-\xi & = & 0.
\label{eq:imaginary part of characteristic equation at edge of stability}
\end{eqnarray}
These equations can be solved for $\tau$  and $\xi$ to yield the critical delay as in eq. (\ref{eq:critical delay without average}).
Note that a critical delay $\tau_c$ only exists, if the price adaptation is larger than the intrinsic damping of the system $\gamma>\alpha$.
Following straight-forward calculations we can prove that eigenvalues obtained from eq. (\ref{eq:simplified characteristic equation of sum}) always destabilize the system, i.e., their real parts are positive for all delays larger than the critical one, 
\begin{equation}
\text{Re}(\lambda(\tau))>0 \quad \forall \tau >\tau_c.
\end{equation}
These results hold for all network topologies, since we needed no assumptions regarding the coupling matrix $K_{ij}$ or the power production $P_i$.
Predicting the precise scaling of the critical delay as a function of the averaging time $T$ is not easily possible but an approximation for small $\xi T$ is obtained as
\begin{equation}
\tau_{c}(T)\approx\frac{\sqrt{T^{2}\gamma^{2}-4}\arctan\left[\frac{\left(\alpha+\gamma\right)\sqrt{T^{2}\gamma^{2}-4}}{\left(2+T\gamma\right)\sqrt{\alpha^{2}-\gamma^{2}}}\right]}{\sqrt{\alpha^{2}-\gamma^{2}}},
\end{equation}
which is a decreasing function in $T$ for parameters $\alpha, \gamma, T>0$. Hence, increasing averaging time $T$ causes the rebound effect to occur for smaller delays $\tau$.

We conclude that the delay $\tau$ has to be smaller than a critical value $\tau_c$ to ensure stability. This critical value depends only on the intrinsic damping $\alpha$ and the price adaptation $\gamma$ and decreases for increasing averaging $T$, while it is valid for all network topologies. Hence, to avoid problems with large delays, we have to enforce all actors of the power grid to react within less then this critical delay $\tau_c$ or need to ensure that intrinsic damping is larger than the price adaptation: $\alpha > \gamma $. For the next section, we restrict ourself to the interval $\tau \in [0,5]\text{s}$ to avoid this destabilizing rebound effect.

\paragraph{Intermediate delays benefit stability.}
\begin{figure}[htb!]
	\centering
	\includegraphics[width=0.7\linewidth]{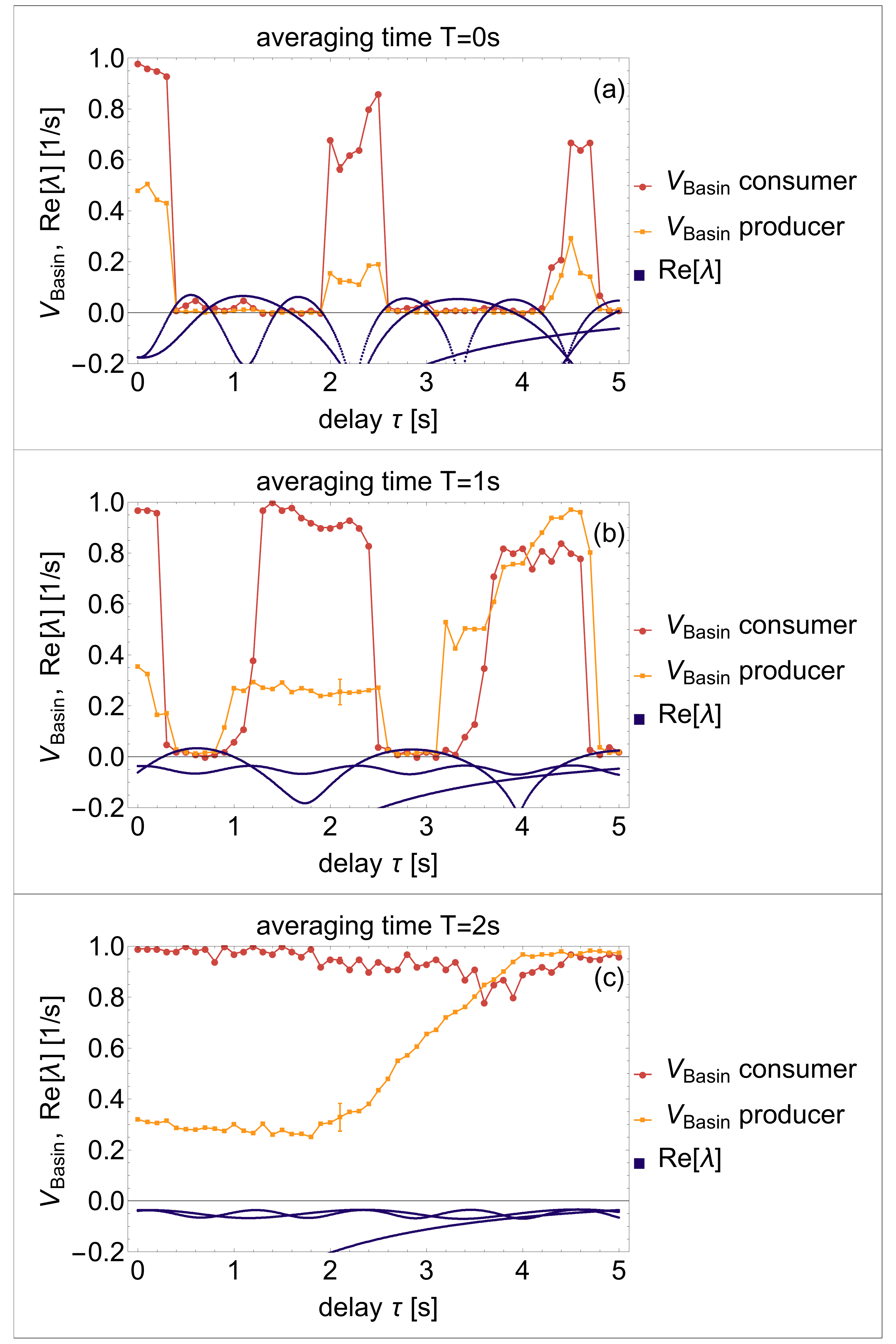}
	\caption{\label{fig:4_node_star_linear_and_basin_Stability}
		\textbf{Stability and basin size for the star topology (see fig. \ref{fig:4_node_eigenvalues}). Intermediate delays result in large basin volume if averaging is switched on.}
		Shown are the real parts of the eigenvalues for the 4 node star motif (dark blue) as well as the basin volume of the producer (dark red) and of one consumer (orange) as functions of the delay $\tau$ for different averaging times: $T_a=0\text{s}$, $T_b=1$s, $T_c=2$s. Parameters $\alpha =0.1/\text{s}$, $K=8/\text{s}^2$ and $\gamma =0.25/\text{s}$ were applied. For delay $\tau=2.1$s simulations were repeated 21 times, averaged and the standard deviation is shown as a typical error bar.}
\end{figure}
With the help of linear stability we observed that delays induce destabilizing resonances which can be suppressed by prosumers responding to averaged frequency data. At the same time large delays destabilize the system by introducing a rebound effect. These results are supplemented by information from basin volume analysis. 
For DSGC with averaging (fig. \ref{fig:4_node_star_linear_and_basin_Stability}b and c), we demonstrate how intermediate delays $\tau$ are beneficial for the stability of the system. The basin volume increases with greater delay  $V_\text{basin}(\tau>0)>V_\text{basin}(\tau=0)$ until, for delays $\tau \approx 4$s, we obtain close to perfect stability with $V_\text{basin}\approx 1$ both for an averaging $T_b=1$s and $T_c=2$s. In the previous paragraph we demonstrated that high averaging times and large delays always destabilize the power grid. 
Hence, we observe a trade-off in curing resonances with averaging and avoiding the rebound effect for delays larger than a critical value $\tau_c$. 
Furthermore, basin volume reveals that disturbances in a consumer node are less likely to destabilize the system than perturbations of the producer (compare dark red and light orange curves in fig. \ref{fig:4_node_star_linear_and_basin_Stability}). This is intuitively clear as there is only one producer and the topology increases its importance even more.

We conclude that Decentral Smart Grid Control can be applied to the star motif if an averaging time of at least $T=2\text{s}$ is used or the price elasticity is smaller than the intrinsic damping $\gamma < \alpha$. Additionally, intermediate delays $\tau \approx 4$s incorporate the trade-off between curing either destabilizing resonances or rebound effects. They increase the basin volume of the system and thereby benefit the overall stability of the power grid.

\subsection{Effect of decentralized production}
\label{Decentralized_production}
In this section we demonstrate that switching from central to decentralized production improves the linear stability in the power grid topologies we investigate for small and intermediate delays. Specifically, we analyze linear stability for moderately sized lattice and cycle networks for different central and decentralized power production.

\begin{figure}[htb]
	\centering
	\includegraphics[width=0.95\linewidth]{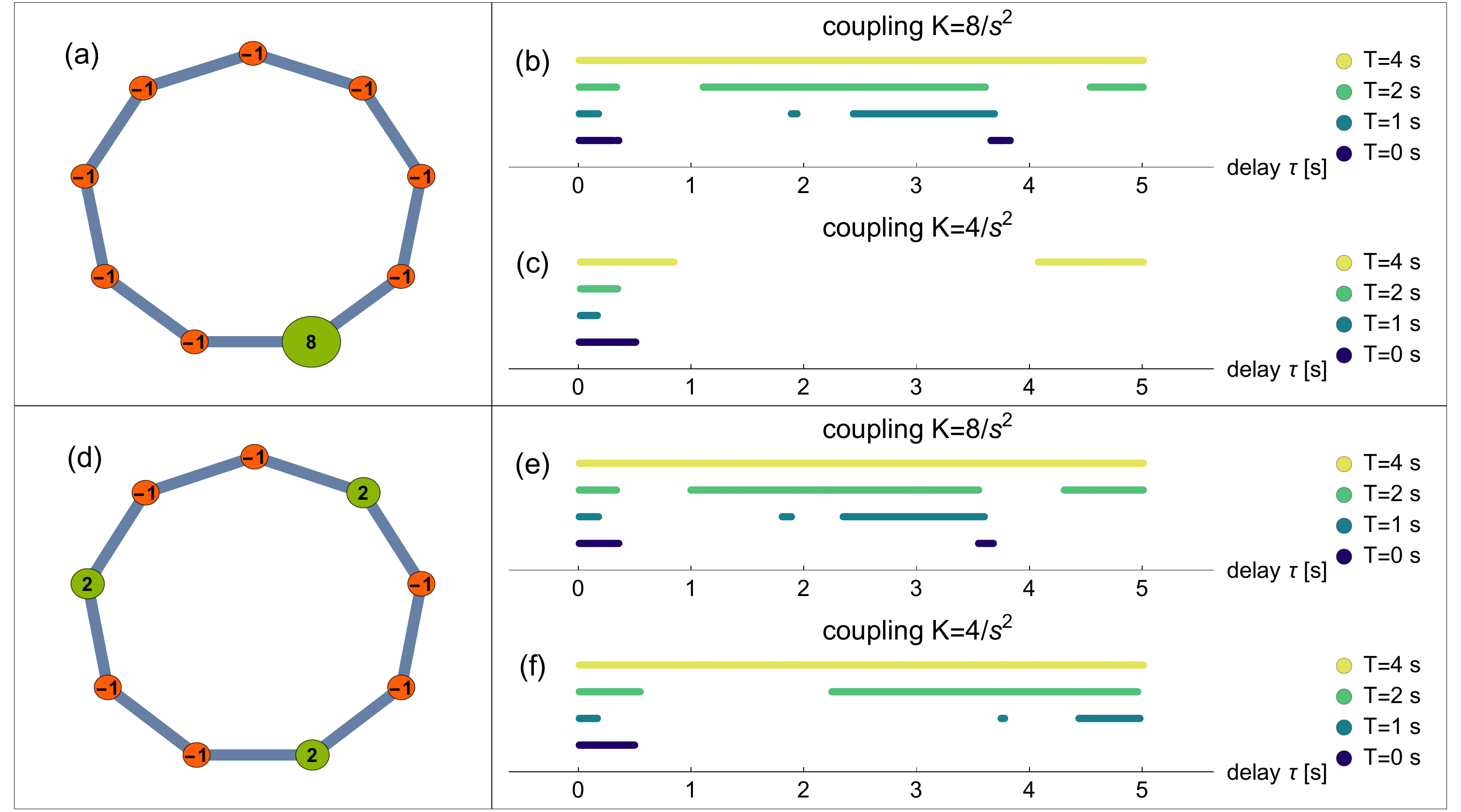}
	\caption{\label{fig:9_node_cycle}
		\textbf{Central power production in a circle network requires larger capacity $K$ than in decentralized power grids.} Shown are the ranges of delay $\tau$ for which the power grid motifs with central production (a) or decentralized production (d) are linearly stable. Panels (b) and (e) present ranges for a high capacity $K=8/\text{s}^2$, whereas (c) and (f) for $K=4/\text{s}^2$. Overall, the regions of stability tend to become larger, the larger the average time $T$. Parameters $\alpha =0.1/\text{s}$ and $\gamma =0.25/\text{s}$ were applied.}
\end{figure}

\begin{figure}[htb]
	\centering
	\includegraphics[width=0.95\linewidth]{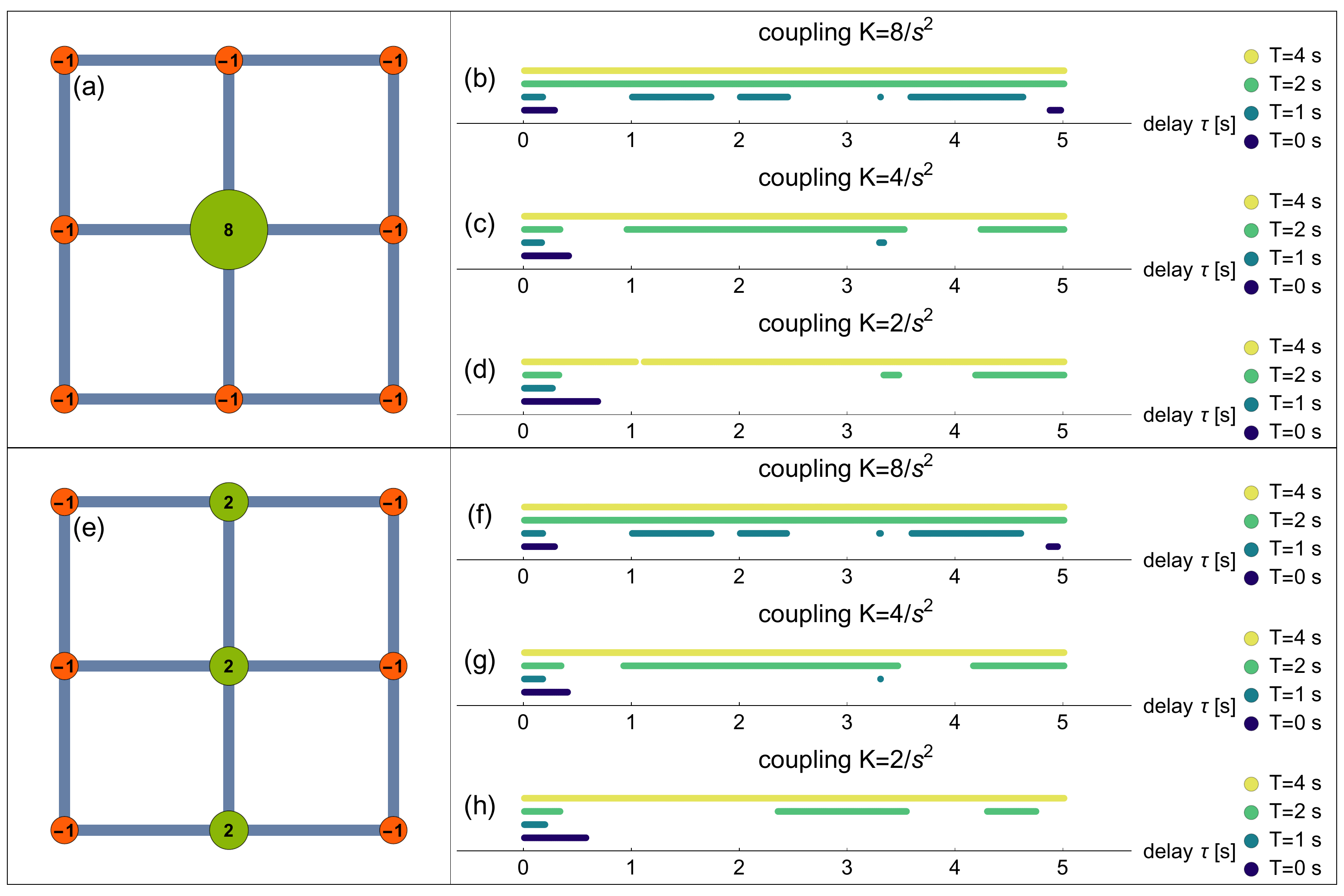}
	\caption{\label{fig:9_node_grid}
		\textbf{Central and decentralized power production in a lattice-like topology lead to similar stability.} In contrast to the cycle network, the lattice-like topology is stable for lower coupling $K$.
		Shown are the ranges of delay $\tau$ for which the power grid motifs with central production (a) or decentralized production (e) are linearly stable. Panels (b) and (f)  present ranges for a high capacity $K=8/\text{s}^2$, (c) and (g) for an intermediate capacity $K=4/\text{s}^2$, finally (d) and (h) for $K=2/\text{s}^2$.
		Overall, the regions of stability tend to become larger, the larger the average time $T$.
		Parameters $\alpha =0.1/\text{s}$ and $\gamma =0.25/\text{s}$ were applied.}
\end{figure}

For a cycle network decentralization enhances stability significantly (fig. \ref{fig:9_node_cycle}). For a power line coupling of $K=8/\text{s}^2$ centralized and decentralized production result in similar stability (fig. \ref{fig:9_node_cycle}b and e). However, when choosing the critical coupling of the cycle network, i.e., the minimal coupling needed so that there exists a fixed point \cite{Rohden2014}, $K=4/\text{s}^2$, the cycle with central production cannot be stabilized for all considered delays, while this is possible for decentralized production (fig. \ref{fig:9_node_cycle}c and f).

A lattice-like topology for power grids allows stable operation with central power production (fig. \ref{fig:9_node_grid}). Choosing large couplings of $K=8/\text{s}^2$ (fig. \ref{fig:9_node_grid}b and e) or even $K=4/\text{s}^2$ (fig. \ref{fig:9_node_grid}c and g), decentralized and centralized production result in very similar stability. Even when operating at the critical coupling of the lattice-like topology $K=2/\text{s}^2$, 
the central power production can be stabilized for sufficiently large averaging time $T=4\text{s}$ (fig. \ref{fig:9_node_grid}d and h).

Note that we chose $\gamma_i=0.25/\text{s}$ for all nodes in the networks. Hence, the large producer with $P_\text{large}=8/\text{s}^2$ adapts relatively less compared to the smaller producers with $P_\text{small}=2/\text{s}^2$. Nevertheless, the overall adaptation of the whole network is 
\begin{equation}
\Delta P=\sum_{i=1}^N \gamma_i \cdot \abs{\Delta \omega},
\end{equation} 
with $\abs{\Delta \omega}$ being the maximal angular frequency deviation.
Hence, the maximal adaptation $\Delta P$ is independent of the power distribution.

We conclude that a centralized power production requires larger transmission capacities compared to a decentralized power production to guarantee stable power grid operation. An averaging time of $T\approx 4\text{s}$ stabilizes the power grid with Decentral Smart Grid Control for all considered delays. Note that our decentralized production utilized short distances to the consumers.
Decentralized power production tends to allow smaller averaging times, thereby offering a greater safe operating space. In addition, a highly connected topology like a lattice outperforms the less connected cycle in terms of stability.

\section{Summary and Discussion}

In this article we applied the concept of "Decentral Smart Grid Control" (DSGC), as proposed in \cite{Schaefer2015}, to different motifs and small networks. We first determined both the linear stability and the basin volume of a 4-node-star motif in dependence on the delay time $\tau$ (see equation \ref{eq.motion with delay}) and for fixed averaging times $T$ (see equation \ref{eq.motion with average}). 
Linear stability analysis reveals two destabilizing effects for the power grid: First, resonance catastrophes destabilize the system periodically. This instability can be cured by applying sufficient averaging (fig. \ref{fig:4_node_eigenvalues}). Secondly, a rebound effect emerges for large delays and destabilizes the system regardless of averaging. The rebound effect sets an upper limit for the delay $\tau=\tau_c$ and magnitude of adaptation response $\gamma$ as it has to be smaller than the intrinsic damping of the system $\alpha$. 
Basin volume analysis gives further probabilistic insight on how well DSGC tames grid instabilities.
For large averaging times $T$ and intermediate delays $\tau$, basin volume approaches unity (fig. \ref{fig:4_node_star_linear_and_basin_Stability}). Hence, for DSGC exists a trade-off in curing resonances with averaging and larger delays and avoiding the rebound effect for delays larger than a critical value $\tau_c$. 

Summarizing the results from linear and basin volume analysis, adaptation has to be smaller than the intrinsic damping of the system ($\gamma<\alpha$) or the demand response time needs to be located in a delay window of safe operation ($\tau<\tau_c$). For values above the critical delay $\tau_c$ the system becomes always destabilized, regardless of the averaging time. At the same time, averaging and increasing delay is beneficial for system stability in terms of basin volume. 
These results have strong implications on how parameters has to be set for real world applications of DSGC.

In the last section of this article, we demonstrated the usefulness of DSGC with centralized as well as with decentralized power production: While it works in both cases, central production requires larger line capacities $K$. For the lattice-like topology, this effect can be compensated by using longer averaging times.
But decentralized power production is clearly advantageous.

Next research steps include considering heterogeneous networks, i.e. the use of different $\tau$, $\gamma$, $T$ values for individual nodes, modifying the averaging method, e.g., to a discrete time window and extending the DSGC framework to larger network topologies. In this context, there remain a couple of open questions that will have to be investigated in more detail, namely:  What is the reason that we observe delays $\tau$ for which $V_\text{basin}(\tau)>V_\text{basin}(\tau=0)$, in particular for larger averaging times? 
Do we need even larger averaging times $T$ when we go to larger networks? How large is the safe operating space to cure instabilities by resonances while avoiding the rebound effect for different networks? These are all widely open questions.

In this article, we have demonstrated that Decentral Smart Grid Control constitutes a promising control concept, in particular for future power grids that will be more decentralized than the present one.

\section*{Acknowledgments}
We thank Dr. Thomas Walter at Easy Smart Grid GmbH for inspiration and valuable discussions.
This work was supported by the BMBF, grants No. 03SF0472A (C.G., S.A., J.K.),No. 03SF0472B (D.W.) and No.
03SF0472E (M.T.), by the Helmholtz Association, grant no.VH-NG-1025 (D.W.), by the Max Planck Society (M.T.), by the Göttingen Graduate School for Neurosciences and Molecular Biosciences (DFG Grant GSC 226/2) (B.S.).

\bibliography{References_revised1}

\begin{thebibliography}{10}

\bibitem{mitigation2011ipcc}
O.~Edenhofer, R.~Madruga, Y.~Sokona, and K.~Seyboth.
\newblock {IPCC} {S}pecial {R}eport on {R}enewable {E}nergy {S}ources and
  {C}limate {C}hange {M}itigation.
\newblock 2011.

\bibitem{EE2014}
Bundesministerium f\"ur Wirtschaft~und Energie.
\newblock {Erneuerbare Energien im Jahr 2014}, 2015.

\bibitem{ackermann2001distributed}
T.~Ackermann, G.~Andersson, and L.~S{\"o}der.
\newblock {Distributed Generation: a Definition}.
\newblock {\em {Electric Power Systems Research}}, 57(3):195--204, 2001.

\bibitem{kotler1986prosumer}
P.~Kotler.
\newblock {The Prosumer Movement: A new Challenge for Marketers}.
\newblock {\em {Advances in Consumer Research}}, 13(1):510--513, 1986.

\bibitem{Turner1999}
J.~A. Turner.
\newblock {A Realizable Renewable Energy Future}.
\newblock {\em Science}, 285(5428):687--689, 1999.

\bibitem{50Hertz2012}
50Hertz, Amprion, TenneT TSO, and TransnetBW.
\newblock {Netzentwicklungplan Strom}, 2012.

\bibitem{Boyle2004}
G.~Boyle.
\newblock {\em {Renewable Energy}}.
\newblock {Oxford University Press, Oxford}, 2004.

\bibitem{Heide2010}
D.~Heide, L.~Von Bremen, M.~Greiner, C.~Hoffmann, M.~Speckmann, and
  S.~Bofinger.
\newblock {Seasonal Optimal Mix of Wind and Solar Power in a Future, Highly
  Renewable Europe}.
\newblock {\em {Renewable Energy}}, 35(11):2483--2489, 2010.

\bibitem{Milan2013}
P.~Milan, M.~W{\"a}chter, and J.~Peinke.
\newblock {Turbulent Character of Wind Energy}.
\newblock {\em {Physical Review Letters}}, 110(13):138701, 2013.

\bibitem{gruenbuch}
{B}undesministerium f\"ur {W}irtschaft und~{E}nergie ({BMW}i).
\newblock Ein {S}trommarkt f\"ur die {E}nergiewende, Oktober 2014.

\bibitem{BEEStudie}
M.~Jansen, C.~Richts, and N.~Gerhardt.
\newblock {Strommarkt-Flexibilisierung - Hemmnisse und L\"osungskonzepte},
  Januar 2015.

\bibitem{Butler2007}
D.~Butler.
\newblock {Energy Efficiency: Super Savers: Meters to manage the Future}.
\newblock {\em Nature}, 445(7128):586--588, 2007.

\bibitem{Albadi2008}
M.~H. Albadi and E.~F. El-Saadany.
\newblock {A Summary of Demand Response in Electricity Markets}.
\newblock {\em {Electric Power Systems Research}}, 78(11):1989--1996, 2008.

\bibitem{Palensky2011}
P.~Palensky and D.~Dietrich.
\newblock {Demand Side Management: Demand Response, Intelligent Energy Systems,
  and Smart Loads}.
\newblock {\em {Industrial Informatics, IEEE Transactions on}}, 7(3):381--388,
  2011.

\bibitem{Kok2005}
J.~K. Kok, C.~J. Warmer, and I.G. Kamphuis.
\newblock {PowerMatcher: Multiagent Control in the Electricity Infrastructure}.
\newblock In {\em {Proceedings of the Fourth International joint Conference on
  Autonomous Agents and Multiagent Systems}}, pages 75--82. ACM, 2005.

\bibitem{Hofmann}
L.~Hofmann and M.~Sonnenschein.
\newblock {Smart Nord Final Report}, 2015.

\bibitem{Ericsson2010}
G.~N. Ericsson.
\newblock {Cyber Security and Power System Communication - Essential Parts of a
  Smart Grid Infrastructure}.
\newblock {\em {Power Delivery, IEEE Transactions on}}, 25(3):1501--1507, 2010.

\bibitem{Fang2012}
X.~Fang, S.~Misra, G.~Xue, and D.~Yang.
\newblock {Smart Grids - The new and improved Power Grid: A Survey}.
\newblock {\em {Communications Surveys \& Tutorials, IEEE}}, 14(4):944--980,
  2012.

\bibitem{meterstudy}
Ernst \&~Young GmbH.
\newblock {Cost-benefit Analysis for the Comprehensive Use of Smart Metering
  Systems - Final Report - Summary}, January 2013.

\bibitem{Schweppe1982}
F.~C. Schweppe.
\newblock {Frequency Adaptive, Power-Energy Re-scheduler}, February~23 1982.
\newblock {US Patent 4,317,049}.

\bibitem{Short2007}
J.~A. Short, D.~G. Infield, and L.~L. Freris.
\newblock {Stabilization of Grid Frequency through Dynamic Demand Control}.
\newblock {\em {Power Systems, IEEE Transactions on}}, 22(3):1284--1293, 2007.

\bibitem{Walter}
T.~Walter.
\newblock {Smart Grid neu gedacht: Ein L\"osungsvorschlag zur Diskussion in
  VDE|ETG}, 2014.

\bibitem{Schaefer2015}
B.~Sch{\"a}fer, M.~Matthiae, M.~Timme, and D.~Witthaut.
\newblock {Decentral Smart Grid Control}.
\newblock {\em {New Journal of Physics}}, 17(1):015002, 2015.

\bibitem{Filatrella2008}
G.~Filatrella, A.~H. Nielsen, and N.~F. Pedersen.
\newblock {Analysis of a Power Grid using a Kuramoto-like Model}.
\newblock {\em {The European Physical Journal B-Condensed Matter and Complex
  Systems}}, 61(4):485--491, 2008.

\bibitem{Witthaut2012}
D.~Witthaut and M.~Timme.
\newblock {Braess's Paradox in Oscillator Networks, Desynchronization and Power
  Outage}.
\newblock {\em {New Journal of Physics}}, 14(8):083036, 2012.

\bibitem{Rohden2014}
M.~Rohden, A.~Sorge, D.~Witthaut, and M.~Timme.
\newblock {Impact of Network Topology on Synchrony of Oscillatory Power Grids}.
\newblock {\em {Chaos: An Interdisciplinary Journal of Nonlinear Science}},
  24(1):013123, 2014.

\bibitem{Doerfler2013}
F.~D{\"o}rfler M., Chertkov F., and Bullo.
\newblock {Synchronization in Complex Oscillator Networks and Smart Grids}.
\newblock {\em {Proceedings of the National Academy of Sciences}},
  110(6):2005--2010, 2013.

\bibitem{Motter2013}
A.~E. Motter, S.~A. Myers, M.~Anghel, and T.~Nishikawa.
\newblock {Spontaneous Synchrony in Power-grid Networks}.
\newblock {\em {Nature Physics}}, 9(3):191--197, 2013.

\bibitem{Menck2014}
P.~J. Menck, J.~Heitzig, J.~Kurths, and H.~J. Schellnhuber.
\newblock {How Dead Ends undermine Power Grid Stability}.
\newblock {\em {Nature Communications}}, 5(3969), 2014.

\bibitem{Rohden2012}
M.~Rohden, A.~Sorge, M.~Timme, and D.~Witthaut.
\newblock {Self-organized Synchronization in Decentralized Power Grids}.
\newblock {\em {Physical Review Letters}}, 109(6):064101, 2012.

\bibitem{Machowski2011}
J.~Machowski, J.~Bialek, and J.~Bumby.
\newblock {\em {Power System Dynamics: Stability and Control}}.
\newblock {John Wiley \& Sons}, 2011.

\bibitem{Bergen1981}
A.~R. Bergen and D.~J. Hill.
\newblock {A Structure preserving Model for Power System Stability Analysis}.
\newblock {\em {Power Apparatus and Systems, IEEE Transactions on}},
  100(1):25--35, 1981.

\bibitem{VanHertem2006}
D.~Van Hertem, J.~Verboomen, K.~Purchala, R.~Belmans, and W.~L. Kling.
\newblock {Usefulness of DC Power Flow for Active Power Flow Analysis with Flow
  Controlling Devices}.
\newblock In {\em {AC and DC Power Transmission, 2006. ACDC 2006. The 8th IEE
  International Conference on}}, pages 58--62. IET, 2006.

\bibitem{Manik2014}
D.~Manik, D.~Witthaut, B.~Sch{\"a}fer, M.~Matthiae, A.~Sorge, M.~Rohden,
  E.~Katifori, and M.~Timme.
\newblock {Supply Networks: Instabilities without Overload}.
\newblock {\em {The European Physical Journal Special Topics}},
  223(12):2527--2547, 2014.

\bibitem{Driver1977}
R.~D. Driver.
\newblock {Introduction to Delay Differential Equations}.
\newblock In {\em {Ordinary and Delay Differential Equations}}, pages 225--267.
  Springer, 1977.

\bibitem{Roussel2004}
M.~R. Roussel.
\newblock {Delay-Differential Equations}, 2005.

\bibitem{WolframResearch2014}
Wolfram~Research Inc.
\newblock Mathematica.
\newblock Champaign, Illinois, 2014.

\bibitem{Gu2003}
K.~Gu, J.~Chen, and V.~L. Kharitonov.
\newblock {\em {Stability of Time-delay Systems}}.
\newblock {Springer Science \& Business Media}, 2003.

\bibitem{Menck2013}
P.~J. Menck, J.~Heitzig, N.~Marwan, and J.~Kurths.
\newblock {How Basin Stability Complements the Linear-stability Paradigm}.
\newblock {\em {Nature Physics}}, 9(2):89--92, 2013.

\bibitem{schultz2014detours}
P.~Schultz and J.~Heitzig andJ. Kurths.
\newblock {Detours around Basin Stability in Power Networks}.
\newblock {\em {New Journal of Physics}}, 16(12):125001, 2014.

\bibitem{ENTSO-E2013}
ENTSO-E.
\newblock Network code on requirements for grid connection applicable to all
  generators (rfg), March 2013.

\bibitem{Naduvathuparambil2002}
Biju Naduvathuparambil, Matthew Valenti, Ali Feliachi, et~al.
\newblock Communication delays in wide-area measurement systems.
\newblock In {\em Southeastern Symposium on System Theory}, volume~34, pages
  118--122. Citeseer, 2002.

\end{thebibliography}
\bibliographystyle{unsrt}

\end{document}